\documentclass[%
 reprint,
nofootinbib,
 amsmath,amssymb,
 aps,
]{revtex4-2}

\usepackage{graphicx}
\usepackage{xcolor}
\usepackage{dcolumn}
\usepackage{bm}
\usepackage{hyperref}
\usepackage{cleveref}
\usepackage{wasysym}

\newcommand{\dd}{\text{d}}
\newcommand{\ii}{\text{i}}
\newcommand{\kB}{k_\textrm{B}}

\begin{document}

\title{The Interplay between Memory and Potentials of Mean Force:\\A Discussion on the Structure of Equations of Motion for Coarse Grained Observables}

\author{Fabian Glatzel}
\email{fabian.glatzel@physik.uni-freiburg.de}
\author{Tanja Schilling}%
\email{tanja.schilling@physik.uni-freiburg.de}
 
\affiliation{%
 Institute of Physics, University of Freiburg, Hermann-Herder-Str. 3, 79104 Freiburg, Germany
}%

\date{\today}

\begin{abstract}
The underdamped, non-linear, generalized Langevin equation is widely used to model coarse-grained dynamics of soft and biological materials. By means of a  projection operator formalism, we show under which approximations this equation can be obtained from the dynamics of the underlying microscopic system and in which cases it makes sense to introduce a potential of mean force.  
We discuss shortcomings of previous derivations presented in the literature and demonstrate the implications of our derivation for the structure of memory terms and for generalized fluctuation-dissipation relations. We show, in particular, that the widely used, simple structure which contains a potential of mean force, a memory term which is linear in the observable, and a fluctuating force which is related to the memory term by a fluctuation-dissipation relation, is neither exact nor can it, in general, be derived as a controlled approximation to the exact dynamics. 
\end{abstract}

\maketitle

When a coarse-grained model for a polymeric system or a biological macromolecule is designed, groups of atoms are merged into larger units. Then approximate equations of motion for these units are solved to predict the evolution of the model \cite{Berendsen07,peter2009,Attinger04}. Depending on context, the coarse-grained degrees of freedom can range from the positions of small molecules or chemical groups to reaction coordinates such as the relative orientation of structural motifs in a biomolecule. In principle, to obtain the exact equation of motion of a coarse-grained degree of freedom, one would need to integrate out systematically the atomistic degrees of freedom. However, as this is a very difficult task, researchers usually resort to effective models. For instance, the underdamped non-linear Langevin equation is frequently used 
\begin{equation}
\label{eq:nlLE} m \ddot{x}_t = -\left.\frac{\dd W(x)}{\dd x} \right\vert_{x=x_t} - \gamma \dot{x}_t + \sqrt{2\gamma \kB T}\xi_t, 
\end{equation}
where $x_t$ is the position of a coarse-grained unit at time $t$ (resp.~the value of a more general reaction coordinate), $m$ is a generalized mass, $\gamma$ is a friction coefficient, $W(x)$ is an effective potential, $\kB T$ is the thermal energy and $\xi_t$ is white Gaussian noise \cite{snook2006}.

The notation $\ddot{x}$ and $\dot{x}$ for the time-derivatives is frequently used in the physics literature to indicate that eq.~\ref{eq:nlLE} could be interpreted in analogy to the Newtonian equation of motion of a particle in a potential energy landscape. However, as $\xi_t$ is a stochastic process, the terms $\ddot{x}$ and $\dot{x}$ are stochastic derivatives, and $W(x)$ is not an external potential but the {\it potential of mean force}
\[W(x) := - k_BT \ln{\left(\rho_X^{\text{eq}}(x)\right)}.\]
Here $\rho_X^{\text{eq}}(x)$ is the so-called ``relevant density'' of the coarse-grained observable $X$, i.e.~the probability of the observable $X$ having the value $x$ in the equilibrium ensemble \cite{grabert2006projection}. Thus the analogy to Newtonian Mechanics can be misleading.
For canonical dynamics one also often encounters the terms {\it effective free energy} and {\it free energy landscape} for $- k_BT \ln{\left(\rho_X^{\text{eq}}(x)\right)}$, denoted by $\Delta G(x)$.

The dynamics of the atomistic degrees of freedom which have been integrated out, in general, produce memory effects. Therefore integro-differential equations are also often used to model coarse-grained variables, such as e.g.
\begin{equation}
\label{eq:nlGLE}
m \ddot{x}_t = - \left. \frac{\dd W(x) }{\dd x}\right\vert_{x=x_t} - \int_0^t \dd  \tau \; K(t-\tau)  \dot{x}_\tau + f_t,
\end{equation}
where  $K$ is the memory kernel and $f$ the flucutating force, which is related to $K$ by the second fluctuation-dissipation theorem
(see ref.~\cite{hernandez1999_2,Bhadauria2015,lei:2016,Daldrop2018,wang2019,Kappler2019,wang2020,Ozmaian2019,grogan2020} for examples of recent work in which this equation is used to model coarse-grained dynamics). In this letter, we type-set times in parentheses if the time-dependence is on the level of the ensemble average (as e.g.~in the memory kernel $K(t-\tau)$) and times as subscripts if the dependence is on the level of the individual trajectory (as e.g.~in $x_t$).

These effective equations of motion are frequently used in the soft matter modelling community. They provide a practical pathway to coarse-grained modelling,  because the functions $W(x)$ and $K(t)$ can be parameterised and then fitted to simulation data. Therefore it is interesting to check under which assumptions these equation can be derived from first principles. We are aware of only two publications, in which derivations for eq.~\ref{eq:nlGLE} are shown \cite{Lange_2006,Kinjo_2007}. Note that we are referring specifically to the form of the generalized Langevin equation, in which the variable $x$ enters $W(x)$ non-linearly, while the memory term is linear in $\dot{x}$, and $K(t-\tau)$ and $f_t$ are related by the second flucutation-dissipation theorem. Other forms of the generalized Langevin equation, as discussed e.g.~in ref.~\cite{snook2006,chorin:2000,chorin2002,Vrugt2020,Hijon_2010,izvekov2017,MeyerVoigtmann_2019}, are not in the focus of our letter. We also acknowledge that, if one replaces the potential of mean force $W(x)$ in eq.~\ref{eq:nlGLE} with an external potential $V_{\rm ext}(x)$, one obtains the equation of motion of a specific, well-known model system: one particle linearly coupled to a bath of harmonic oscillators \cite{zwanzig2001nonequilibrium}. Also this is not the type of problem we are referring to in this letter. We are aiming at coarse-graining the dynamics of complex systems such as polymers and biomolecules, in and out of equilibrium.
The question we address in this letter is, if an equation of motion with the structure of \cref{eq:nlGLE} can be derived for a broad class of systems and observables, as claimed in ref.~\cite{Lange_2006} and \cite{Kinjo_2007}.

A useful framework to  tackle this task is the projection operator formalism as originally introduced by Zwanzig \cite{Zwanzig_60,Zwanzig_61} and Mori \cite{Mori_65}.
Let $\Gamma = \left\{q_i,p_i\right\}$ denote the phase space coordinates of the microscopic system and $\ii\mathcal{L}$ the Liouvillian, which for now shall not be explicitly time-dependent (we will come to time-dependent Liouvillians later). $\vec{\mathbb{A}}(\Gamma)$ shall denote a set of phase space fields, e.g.~the coarse grained observables for which we intend to derive an equation of motion. We use blackboard-bold for phase-space functions and italics for the value they take at specific points in phase space, i.e.~$\vec{A}_t=\vec{\mathbb{A}}(\Gamma_t)$. In the case of Hamiltonian dynamics, the action of the Liouvillian on the fields $\mathbb{A}_i(\Gamma)$ is given by 
\begin{align}
	\dot{\mathbb
	A}_i&=\ii\mathcal{L}\mathbb{A}_i = \left\{\mathbb{A}_i,\mathbb{H}\right\}=\sum\limits_j\frac{\partial \mathbb{A}_i}{\partial q_j}\frac{\partial\mathbb{H}}{\partial p_j}-\frac{\partial \mathbb{A}_i}{\partial p_j}\frac{\partial\mathbb{H}}{\partial q_j}.
\end{align}
The equation of motion of the observables can be integrated formally
\begin{align}
	\frac{\dd \vec{A}_t}{\dd t} &= \exp(t\ii\mathcal{L})\ii\mathcal{L}\vec{\mathbb{A}}(\Gamma)\big|_{\Gamma=\Gamma_0}.\label{eq:LiouvillePropagation}
\end{align}
The right-hand side is the time-evolution operator for a time span of $t$ acting on the time-derivative of $\vec{\mathbb{A}}$. (Note that the initial phase space coordinates $\Gamma_0$ are inserted after performing this operation.)
We introduce projection operators $\mathcal{P}$ which act on the space of phase space fields.
Using the Dyson-Duhamel identity,  \cref{eq:LiouvillePropagation} can be written as
	\begin{align}\label{eq:generalProjectionEquation}
		\frac{\dd \vec{A}_t}{\dd t} &= \int\limits_0^t\dd s\,\exp(s\ii\mathcal{L})\mathcal{P}\ii\mathcal{L}\mathcal{Q}\exp\left((t-s)\ii\mathcal{L}\mathcal{Q}\right)\ii\mathcal{L}\vec{\mathbb{A}}\nonumber\\
		&\phantom{=}+\exp(t\ii\mathcal{L})\mathcal{P}\ii\mathcal{L}\vec{\mathbb{A}}+\mathcal{Q}\exp\left(t\ii\mathcal{L}\mathcal{Q}\right)\ii\mathcal{L}\vec{\mathbb{A}} ,
	\end{align}
 where $\mathcal{Q} := 1-\mathcal{P}$.
(We dropped the explicit insertion of the initial point in phase space $\Gamma_0$ on the right-hand side.) 
Next, we need to choose a specific projector. Two of the most prominent types of projectors are the Zwanzig and the Mori projector. As we will see in the following, the Zwanzig projector has the useful property that the second term in \cref{eq:generalProjectionEquation} turns into a derivative of a potential of mean force under certain conditions. On the other hand, the Mori projector, which is linear in the observable(s), yields a fluctuation-dissipation relation.

We begin with a projection operator similar to Zwanzig's original one to bring the second term of the right-hand side of \cref{eq:generalProjectionEquation} into the form of a derivative of a potential of mean force.
We define
\begin{align}\label{eq:definitionProjector}
		\mathcal{P}\mathbb{X}(\Gamma) &= \frac{1}{\rho_A(\vec{\mathbb{A}}(\Gamma))}\int\dd\Gamma'\rho^{\text{eq}}(\Gamma')\mathbb{X}(\Gamma')
		\delta(\vec{\mathbb{A}}(\Gamma')-\vec{\mathbb{A}}(\Gamma))\\
		\intertext{with}
			\rho_A(\vec{A}) &= \int\dd\Gamma\,\rho^{\text{eq}}(\Gamma)\delta(\vec{\mathbb{A}}(\Gamma)-\vec{A}),
\end{align}
where $\mathbb{X}$ is an arbitrary phase space field.
The normalization factor $\rho_A(\vec{A})$ is the relevant density of the variables $\mathbb{A}_i$. In the next steps, we use the canonical equilibrium density $\rho^\text{eq}(\Gamma)\propto\exp(-\beta\mathbb{H}(\Gamma))$, where $\beta=1/\kB T$, to project $\mathbb{X}$ onto a set of phase space fields $\{\mathbb{A}_i\}$. (This is not mandatory, a similar derivation can also be carried out for other ensembles.)

To derive an equation of a structure similar to eq.~\ref{eq:nlGLE}, we assume that the Hamiltonian can be split into a simple kinetic contribution and a potential that depends only on the generalized coordinates
\begin{align}
	\mathbb{H}(\Gamma) &= \sum\limits_{j} \frac{p_j^2}{2m_j} + \mathbb{V}(q_1,\cdots,q_N).\label{eq:Hamiltonian}
\end{align}
Now we project onto a set of two rather specific observables, $\vec{\mathbb{A}}=(\mathbb{A}_1,\mathbb{A}_2)^\top$.
The first one takes the form $\mathbb{A}_1=\alpha_0+\sum_j\alpha_jq_j$ with constant prefactors $\alpha_i$. An example for such an observable would be a center of mass of a set of atoms.
The second observable is the time derivative of the first one, namely $\mathbb{A}_2 = \ii\mathcal{L}\mathbb{A}_1 = \sum_j \alpha_jp_j/m_j$.\\
To obtain an equation similar to \cref{eq:nlGLE}, where the left-hand side is the second time-derivative of the observable, we consider the evolution of $\mathbb{A}_2$. Thus, the second term in the second component of \cref{eq:generalProjectionEquation} reads
\begin{align}
	\exp(t\ii\mathcal{L})\mathcal{P}\ii\mathcal{L}\mathbb{A}_2 &= -\exp(t\ii\mathcal{L})\frac{1}{\rho_A(\vec{\mathbb{A}}(\Gamma))}\int\dd\Gamma'\,\rho^\text{eq}(\Gamma')\nonumber\\
	&\phantom{=}\times\delta(\vec{\mathbb{A}}\left(\Gamma'\right)-\vec{\mathbb{A}}(\Gamma))\sum\limits_{j}\frac{\alpha_j}{m_j}\frac{\partial \mathbb{V}(\Gamma')}{\partial q_j'}.\label{eq:projectorEquationFirstTerm}
\end{align}
To relate this expression to a potential of mean force, we take the derivative of the relevant density with respect to the first coordinate
\begin{align}
	\frac{\partial \rho_A\left(\vec{A}\right)}{\partial A_1}
	&= -\int\dd\Gamma'\,\rho^\text{eq}\left(\Gamma'\right)\delta\left(\mathbb{A}_2\left(\Gamma'\right)-A_2\right)\nonumber\\
	&\phantom{=}\times\mu\sum\limits_j\frac{\alpha_j}{m_j}\frac{\partial}{\partial q_j'}\delta\left(\mathbb{A}_1\left(\Gamma'\right)-A_1\right)
\end{align}
where $\mu=\sum_i m_i/\alpha_i^2$. Next, we carry out an integration by parts and use the fact that
\begin{align}
	\frac{\partial }{\partial q_j}\rho^\text{eq}(\Gamma) &= -\beta \frac{\partial\mathbb{V}(\Gamma)}{\partial q_j}\rho^\text{eq}(\Gamma)
\end{align}
to obtain
\begin{align}
	\frac{\partial \rho_A(\vec{A})}{\partial A_1} &= -\beta\mu\int\dd\Gamma'\,\rho^\text{eq}\left(\Gamma'\right)\delta(\vec{\mathbb{A}}\left(\Gamma'\right)-\vec{A})\nonumber\\
	&\phantom{=}\times \sum\limits_j \frac{\alpha_j}{m_j}
	\frac{\partial\mathbb{V}(\Gamma')}{\partial q_j'}\label{eq:derivativeRelevantDensity}.
\end{align}
Comparing \cref{eq:projectorEquationFirstTerm} and \cref{eq:derivativeRelevantDensity}, we see that
\begin{align}
	\exp(t\ii\mathcal{L})\mathcal{P}\ii\mathcal{L}\mathbb{A}_2 &= -\frac{1}{\mu}\frac{\partial}{\partial A_1}\underbrace{\left(-\kB T\ln(\rho_A(\vec{A}))\right)}_{=:W(\vec{A})}\Big|_{\vec{A}_t}\quad ,
\end{align}
where $W(\vec{A})$ is the potential of mean force. Note that $\partial W(\vec{A})/\partial A_1$ does not depend on $A_2$, because the $A_2$-dependent factors in the numerator and denominator cancel each other.

A similar expression can be derived for a projection on multiple observables, e.g.~for a projection on the centers of mass of several ``blobs'' (united atoms in a polymeric system). In this case one replaces the first observable $\mathbb{A}_1$ by a set of observables $\mathbb{A}_{1i}$, where each of these observables has the form $\mathbb{A}_{1i}=\alpha_{0i}+\sum_j \alpha_{ji}q_j$ with constant prefactors $\alpha_{ji}$. Then the set is extended by the time-derivatives $\mathbb{A}_{2i}=\ii\mathcal{L}\mathbb{A}_{1i}$. Similar as before, the effective masses $\mu_i=\sum_jm_j/\alpha^2_{ji}$ are defined. However, there is one additional restriction in this case: If a microscopic coordinate $q_j$ enters one $A_{1k}$ it must not enter any other $A_{1l}$. Otherwise, the partial integration performed to obtain \cref{eq:derivativeRelevantDensity} will yield additional terms. If the coarse-grained observables are the centers of mass of different blobs, this means that a single particle must not be attributed to more than one blob.\\
In this case the derivation can be carried out as before and we obtain
\begin{align}
	\exp(t\ii\mathcal{L})\mathcal{P}\ii\mathcal{L}\mathbb{A}_{2i} &= -\frac{1}{\mu_i}\frac{\partial}{\partial A_{1i}}\underbrace{\left(-\kB T\ln(\rho_A(\vec{A}))\right)}_{=:W(\vec{A})}\Big|_{\vec{A}_t}\quad .
\end{align}
Again, $\partial W(\vec{A})/\partial A_{1i}$ does not depend on the $A_{2i}$.

Thus, the second term in \cref{eq:generalProjectionEquation} takes the form of a derivative of a potential of mean force under the conditions
\begin{itemize}
	\item[-] The Hamiltonian is of the form given in \cref{eq:Hamiltonian}.
	\item[-] A Zwanzig-type projector (cf.~\cref{eq:definitionProjector}) onto two observables is used.
	\item[-] The first set of observables of the projector is of the form $\mathbb{A}_{1i}=\alpha_{0i}+\sum_j \alpha_{ji}q_j$ with constant $\alpha_{ji}$.
	\item[-] If a coordinate $q_j$ enters one $A_{1k}$ it must not enter any other $A_{1l}$.
	\item[-] The second set of observables of the projector is the time-derivative of the first set $\mathbb{A}_{2i}=\ii\mathcal{L}\mathbb{A}_{1i}$.
\end{itemize}
(We note that this derivation does also hold for a time-dependent Hamiltonian if it can be expressed in the form of \cref{eq:Hamiltonian} with time-dependent masses and/or a time-dependent potential $\mathbb{V}(q_1,\cdots,q_N;t)$. In this case, we would need a time-dependent projector where the equilibrium density in \cref{eq:definitionProjector} is replaced by the equilibrium density with respect to the current Hamiltonian $\rho^\text{eq}(\Gamma;t)\propto\exp(-\beta\mathbb{H}(\Gamma;t))$. See ref.~\cite{MeyerVoigtmann_2019} for a suitable projection operator approach.)

Next, we consider the first and third term of \cref{eq:generalProjectionEquation}. Inserting the Zwanzig projector, \cref{eq:definitionProjector}, into the first term of the right-hand side of \cref{eq:generalProjectionEquation} we obtain a term which is in general nonlinear in $\dot{A}_\tau$. Based on Zwanzig's work \cite{Zwanzig_61}, Hijon et al.~showed in ref.~\cite{Hijon_2010} 
that the memory term for the second component can be written as
\begin{align*}
          \int\limits_0^t\dd s\!\sum_{i=1,2}\! \left(M_i(\vec{A}_s,t-s)\frac{\partial W(\vec{A}_s)}{\partial A_{i,s}}+\kB T \frac{\partial M_i(\vec{A}_s,t-s)}{\partial A_{i,s}}\right)
\end{align*}
with
\[
  M_i(\vec{A},t) := \frac{1}{\kB T}\mathcal{P}\left(\left[\ii\mathcal{L}\mathbb{A}_i\right]\left[\mathcal{Q}\exp(t\ii\mathcal{L}\mathcal{Q})\ii\mathcal{L}\mathbb{A}_2\right]\right).
\]
(The sum runs over the number of components of $\vec{\mathbb{A}}$.)
This expression differs considerably from the desired one in \cref{eq:nlGLE}. Even if we assume that there is time-scale separation between the observables $\mathbb{A}_i$ and the other degrees of freedom, such that $M_i(\vec{A}_s,t-s)\propto \gamma(\vec{A}_s)\delta(t-s)$, we do not recover eq.~\ref{eq:nlLE}, because in general, $\gamma(\vec{A}_s)$ is not a constant.

In order to see the structure of the first and third term of \cref{eq:generalProjectionEquation} more clearly, we now use a mapping of the Zwanzig projector to a Mori projector as proposed in ref.~\cite{Kawai_2011,Izvekov}. (This mapping holds for a general set $\{\mathbb{A}_i\}$, not just for the specific one used in the previous paragraphs.)
 We define a scalar product between square-integrable phase space fields by
\begin{align}
	\left(\mathbb{X},\mathbb{Y}\right) &= \int\dd\Gamma\,\rho^{\rm eq}(\Gamma)\mathbb{X}(\Gamma)\mathbb{Y}(\Gamma),\label{eq:defScalarProductEquilibrium}
\end{align}
and express the projector in \cref{eq:definitionProjector} as
\begin{align}
	\mathcal{P}\mathbb{X}(\Gamma) &= \left.\frac{(\delta(\vec{\mathbb{A}}-\vec{A}),\mathbb{X})}{(\delta(\vec{\mathbb{A}}-\vec{A}),1)}\right|_{\vec{A}=\vec{\mathbb{A}}(\Gamma)}.\label{eq:projectorScalarProductForm}
\end{align}
The set of phase space functions that depend on $\Gamma$ solely through $\vec{\mathbb{A}}(\Gamma)$ are a closed subset of all phase space functions. Thus, we can define a complete (infinite) set of phase space functions $f_i(\vec{\mathbb{A}}(\Gamma))$ such that
\begin{align}
	(f_i(\vec{\mathbb{A}}),f_j(\vec{\mathbb{A}})) &= \delta_{i,j}
\end{align}
and
\begin{align}
	\sum\limits_{i=1}^\infty f_i(\vec{\mathbb{A}}(\Gamma))f_i(\vec{\mathbb{A}}(\Gamma')) &= \frac{\delta(\vec{\mathbb{A}}(\Gamma)-\vec{\mathbb{A}}(\Gamma'))}{(\delta(\vec{\mathbb{A}}-\vec{A}),1)\big|_{\vec{A}=\vec{\mathbb{A}}(\Gamma)}}.\label{eq:completeSetBasisFunctions}
\end{align}
These functions form a basis for the subspace of phase space functions that depend on $\Gamma$ solely through $\mathbb{A}_i$. In practice, such a set of basis functions can be obtained by means of a Gram-Schmidt process starting from monomials in $\mathbb{A}_i$. Note, that the denominator in \cref{eq:completeSetBasisFunctions} is a mere consequence of the normalization of the basis functions. Thus, we can write the Zwanzig projector in \cref{eq:projectorScalarProductForm} as
\begin{align}
  \label{eq:ZwanzigAsMori}
	\mathcal{P}\mathbb{X}(\Gamma) &= \sum\limits_{i=1}^\infty (f_i(\vec{\mathbb{A}}),\mathbb{X})f_i(\vec{\mathbb{A}}(\Gamma)).
\end{align}
However, this is nothing but a Mori projector on the infinitely many observables $f_i(\vec{\mathbb{A}})$. Using this expression, we can write the second term in \cref{eq:generalProjectionEquation} as
	\[
		\sum\limits_{i=1}^\infty\int\limits_0^t\dd s\, K_{2,i}(t-s)f_i(\vec{A}_s)\]
		with
		\[K_{j,i}(t-s) = \left(f_i\left(\vec{\mathbb{A}}\right),\ii\mathcal{L}\mathcal{Q}\exp\left((t-s)\ii\mathcal{L}\mathcal{Q}\right)\ii\mathcal{L}\mathbb{A}_j\right).
\]
Using the shorthand notation 
\begin{align}
	\vec{\varepsilon}_t &= \mathcal{Q}\exp\left(t\ii\mathcal{L}\mathcal{Q}\right)\ii\mathcal{L}\vec{\mathbb{A}}.\label{eq:defFullyOrthogonalPart}
\end{align}
we obtain the equation of motion
\begin{align}
	\frac{\dd^2A_{1,t}}{\dd t^2}&=\frac{\dd A_{2,t}}{\dd t}\nonumber \\
	&=\sum\limits_{i=1}^\infty\int\limits_0^t\dd s\, K_{2,i}(t-s)f_i\left(\vec{A}_s\right)\nonumber\\
	&\phantom{=}-\frac{1}{\mu}\frac{\partial}{\partial A_1}W(A_1)\Big|_{A_1=A_{1,t}}+\varepsilon_{2,t}.\label{eq:exactEOMEquilibrium}
\end{align}
If, in particular, $\mathbb{A}$ is the center of mass position of a set of atoms eq.~\ref{eq:exactEOMEquilibrium} reads
\[
	\ddot{x}_t=-\frac{1}{\mu}\frac{\partial}{\partial x}W(x)\Big|_{x=x_t}+\sum\limits_{i=1}^\infty\int\limits_0^t\dd s\, K_{2,i}(t-s)f_i\left(x_s,\dot{x}_s\right)
+\varepsilon_{2,t},
\]
which differs considerably from eq.~\ref{eq:nlGLE}.

As the Liouvillian is anti-self-adjoint
we can write the memory kernels as
\begin{align}
	K_{j,i}(t) &= -(\ii\mathcal{L}f_i(\vec{\mathbb{A}}),\mathcal{Q}\exp\left(t\ii\mathcal{L}\mathcal{Q}\right)\ii\mathcal{L}\mathbb{A}_j).
\end{align}
If the phase space distribution at time $t$ equals $\rho^\text{eq}$, the scalar product can be interpreted as a correlation,
	\begin{align}
		K_{j,i}(t) &= -\left\langle \left(\left.\frac{\dd f_i(\vec{A}_t)}{\dd t}\right|_{t=0}\right)\varepsilon_{j,t}  \right\rangle\nonumber\\
		&= -\left\langle \left(\vec{\varepsilon}_0\cdot\left.\frac{\dd f_i(\vec{A})}{\dd \vec{A}}\right|_{\vec{A}=\vec{\mathbb{A}}(\Gamma_0)}\right)\varepsilon_{j,t} \right\rangle,
	\end{align}
        i.e.~there is a relation between correlations in the fluctuating force and the memory kernel, but all functions $f_i(A_t)$ enter this relation.
This is as close as we get to a fluctuation-dissipation relation. Thus we conclude, if we enforce the drift term in the equation of motion to be a derivative of a potential of mean force, then the memory term and the fluctuating force term are not related by a fluctuation-dissipation relation. 

Via the functions $f_i$ the memory term of \cref{eq:exactEOMEquilibrium} contains all powers and combinations of the variables $\mathbb{A}_i$, not just linear terms. To obtain an expression which is closer in structure to eq.~\ref{eq:nlGLE} (i.e.~one in which the integrand is linear in the observable), we begin the Gram-Schmidt procedure with the linear term $\mathbb{A}_2$ and ensure $f_1(\vec{A}_t)=\phi A_{2,t}$ where $\phi$ is the normalization factor. Then eq.~\ref{eq:nlGLE} follows if we truncate the sum in \cref{eq:exactEOMEquilibrium} at $i=1$. However, this sum is in general not an expansion in a small parameter and, hence, we have no information on the magnitude of the other terms. Thus they should not be dropped without verifying that this constitutes a reasonable approximation for the specific system at hand.

In the case where the fluctuations of the coarse-grained variables around their means are small, Kauzlaric et al. showed that the Zwanzig projector can be approximated by a ``Mori-like'' projector \cite{Kauzlaric2011}. Under these conditions, the memory term can be approximated as one that is linear in the observable. However, the same applies to the drift term, therefore one then does not obtain an equation of motion that contains a non-linear generalized drift (which would allow to define a potential of mean force). A special exception to this is the case where the potential of mean force is quadratic such that its derivative is linear and coincides with the Mori drift term. However, this is certainly not the general case.

We conclude eq.~\ref{eq:nlGLE} is neither exact nor, in general the result of a controlled approximation. The authors of ref.~\cite{Lange_2006} came to a different conclusion, because they switched between a Zwanzig projector and a Mori projector for a single variable (rather than the inifinitely many variables needed for eq.~\ref{eq:ZwanzigAsMori}) half-way through their derivation. In the work of Kinjo et al., the time-dependences in eq.~(17) of ref.~\cite{Kinjo_2007} and eq.~(26) of ref.~\cite{Kinjo_2007} do not match up and, hence, the projector is implicitly switched as well. Unfortunately, in eq.~(26) of ref.~\cite{Kinjo_2007} the time-dependences are not given expicitly.

Finally, we note that $\vec{\varepsilon}_t$, \cref{eq:defFullyOrthogonalPart}, is orthogonal to any phase space field $g(\vec{\mathbb{A}}(\Gamma))$ which depends on the phase space coordinates solely through $\vec{\mathbb{A}}$. Thus, if the phase space distribution equals $\rho^\text{eq}$ at all times, 
\begin{subequations}
	\begin{align}
		\left\langle g_0 \vec{\varepsilon}_t\right\rangle &= \vec{0}\qquad\forall t.
	\end{align}
\end{subequations}
Note, that also $\left\langle\vec{\varepsilon}_t\right\rangle=0\;\forall t$ as we could choose $g(\vec{\mathbb{A}})=\text{const}$.

 Now we extend the discussion to full non-equilibrium, i.e.~we allow for an explicit time-dependence of the Liouvillian. To simplify the projection operator formalism, we ``augment'' phase space by one additional dimension (time) \cite{Kawai_2011,MeyerVoigtmann_2019}. The new coordinates are $\Gamma^{\rm a} = (\Gamma,\tau)$, where the superscript a stands for ``augmented phase space''.
We denote observable fields on the augmented phase space by $\mathbb{A}^{\rm a}(\Gamma^{\rm a})$. However, $\mathbb
{A}^\text{a}$ shall not depend on $\tau$ explicitly. The equivalent to the Liouville operator is
\[
\ii\mathcal{L}^{\rm a}\circ = \dot{\Gamma}^{\rm a}(\Gamma^{\rm a})\cdot\frac{\partial}{\partial \Gamma^{\rm a}}\circ
  \] and observables  evolve according to the equation
\begin{equation}
  \label{eq:EOMAugmented}
A_t(\Gamma^{\rm a}) = \exp\left(\ii\mathcal{L}^{\rm a}t\right)\mathbb{A}^{\rm a}(\Gamma^a).
\end{equation}
We introduce an inner product on the augmented phase space
\begin{widetext}
\[
\left(\mathbb{X}^{\rm a},\mathbb{Y}^{\rm a}\right)^{\rm a}_t = \int\dd \Gamma^{\rm a} \;\rho^{\rm a}(\Gamma^{\rm a},t)\; \mathbb{X^{\rm a}}(\Gamma^{\rm a})\mathbb{Y^{\rm a}}(\Gamma^{\rm a}) = \int \dd \Gamma^{\rm a} \;\rho^{\rm a}(\Gamma^{\rm a},0) \left(\exp\left(\ii\mathcal{L}^{\rm a}t\right)\mathbb{X}^{\rm a}(\Gamma^{\rm a})\right)\;\left(\exp\left(\ii\mathcal{L}^{\rm a}t\right)\mathbb{Y}^{\rm a}(\Gamma^{\rm a})\right),
\]
\end{widetext}
where the notation  $\rho^{\rm a}(\Gamma^{\rm a},t)$ indicates that we synchronized the phase space distribution such that $\rho^{\rm a}(\Gamma^{\rm a},t)= \rho(\Gamma,t)\delta(\tau-t)$. As above, we define an orthonormal basis $\{\phi^{\rm a}_i (\mathbb{A}^{\rm a},\tau)\}$ such that
\[
\left(\phi^{\rm a}_i (\mathbb{A}^{\rm a},\tau),\phi^{\rm a}_j (\mathbb{A}^{\rm a},\tau)\right)_t=\delta_{i,j} \quad \forall t.
\]
Note that we will, in general, need a different set of basis functions for each time $t$.
As a generalized version of the Zwanzig projector, we define
\begin{align*}
\mathcal{P}(t)\mathbb{X}^\text{a}(\Gamma^\text{a})\! =\! \frac{\int \dd {\Gamma^{\rm a}}' \rho^\text{a}({\Gamma^\text{a}}',t) \delta(\mathbb{A}^\text{a}({\Gamma^\text{a}}')\!-\!\mathbb{A}^\text{a}(\Gamma^\text{a}))\mathbb{X}^\text{a}({\Gamma^\text{a}}')}{\int \dd {\Gamma^\text{a}}' \; \rho^\text{a}({\Gamma^\text{a}}',t) \delta(\mathbb{A}^\text{a}({\Gamma^\text{a}}')-\mathbb{A}^\text{a}(\Gamma^\text{a}))}.
\end{align*}
Using the basis set, this projector can be brought into the form
\begin{align*}
\mathcal{P}(t)\mathbb{X}^{\rm a}(\Gamma^{\rm a}) = \sum_{i=1}^\infty
(\phi^{\rm a}_i (\mathbb{A}^{\rm a}(\Gamma^{\rm a}),\tau), \mathbb{X}^{\rm a}({\Gamma^{\rm a}}'))_t\, \phi^{\rm a}_i (\mathbb{A}^{\rm a}(\Gamma^{\rm a}),t).
\end{align*}
In contrast to eq.~\ref{eq:ZwanzigAsMori}, this expression is not a Mori projector on the augmented space. However, as it is linear in the functions $\phi^a_i$, it can still be inserted straight-forwardly into the Dyson-Duhamel identity. We obtain the equation of motion
\begin{equation}\label{eq:nsnlGLE}
\frac{\dd A_t}{\dd t} = \sum_{i=0}^\infty\left(\omega_i(t)\phi^{\rm a}_i (A_t,t) + \int_{0}^t\dd s\,K_i(t,s)\phi^{\rm a}_i (A_s,s)\right) + f_{t},
\end{equation}
with
\begin{align*}
	&\omega_i(t) = (\phi^{\rm a}_i(\mathbb{A}^{\rm a}({\Gamma^{\rm a}}),\tau),\ii\mathcal{L}^{\rm a}\mathbb{A}^{\rm a}({\Gamma^{\rm a}}))_t,\\
	&K_i(t_1,t_2) = \\
	&\quad(\phi^{\rm a}_i(\mathbb{A}^{\rm a}({\Gamma^{\rm a}}),\tau),(\ii\mathcal{L}^{\rm a}\!-\!\dot{\mathcal{P}}(t_2))\mathcal{Q}(t_2) \mathcal{G}(t_2,t_1)\ii\mathcal{L}^{\rm a}\mathbb{A}^{\rm a}({\Gamma^{\rm a}}'))_{t_2},
\end{align*}
and
\begin{align}
	f_{t}=\mathcal{Q}(0)\mathcal{G}(0,t)\ii\mathcal{L}^{\rm a}\mathbb{A}^{\rm a}({\Gamma^{\rm a}}').
\end{align}
where $\mathcal{Q}(t) = 1-\mathcal{P}(t)$ and the negatively time-ordered exponential $\mathcal{G}(t_1,t_2)=\exp_-(\int_{t_1}^{t_2}\dd s\,\ii\mathcal{L}^a\mathcal{Q}(s))$.

If we again impose the condition, that $\phi_1^{\rm a}(\mathbb{A}^{\rm a})\propto \mathbb{A}^{\rm a}$, we could, in principle, truncate the sum in eq.~\ref{eq:nsnlGLE} at $i=1$ in order to obtain a memory term linear in $A_s$. However, as above this sum is not an expansion in a small parameter thus the truncation does not produce a well-controlled approximation.

In summary, we discussed the structure of the non-linear, generalized Langevin equation for a set of coarse-grained observables. By means of a projection operator formalism, we showed that the widely used equation of motion, which consists of a derivative of a potential of mean force as the organized drift, a memory term which is linear in the observable and a fluctuating force which obeys a fluctutation-dissipation relation with respect to the memory kernel, is in general not exact.

To expand the memory kernel in a set of orthogonal polynomials and to then truncate this expansion after the linear contribution does not constitute a general pathway to a controlled approximation to the exact dynamics. Whether or not the combinaton of a potential of mean force with a linear memory term serves as a suitable approximation to a system's coarse-grained dynamics, therefore needs to be tested case by case.

\section{Acknowledgments}
The authors acknowledge funding by the Deutsche Forschungsgemeinschaft (DFG, German Research Foundation)—Project No.~430195928 and No.~431945604 (project P4 in FOR 5099). Further, the authors thank the Erwin Schr\"odinger Institute (ESI).

%

\begin{thebibliography}{29}%
	\makeatletter
	\providecommand \@ifxundefined [1]{%
		\@ifx{#1\undefined}
	}%
	\providecommand \@ifnum [1]{%
		\ifnum #1\expandafter \@firstoftwo
		\else \expandafter \@secondoftwo
		\fi
	}%
	\providecommand \@ifx [1]{%
		\ifx #1\expandafter \@firstoftwo
		\else \expandafter \@secondoftwo
		\fi
	}%
	\providecommand \natexlab [1]{#1}%
	\providecommand \enquote  [1]{``#1''}%
	\providecommand \bibnamefont  [1]{#1}%
	\providecommand \bibfnamefont [1]{#1}%
	\providecommand \citenamefont [1]{#1}%
	\providecommand \href@noop [0]{\@secondoftwo}%
	\providecommand \href [0]{\begingroup \@sanitize@url \@href}%
	\providecommand \@href[1]{\@@startlink{#1}\@@href}%
	\providecommand \@@href[1]{\endgroup#1\@@endlink}%
	\providecommand \@sanitize@url [0]{\catcode `\\12\catcode `\$12\catcode
		`\&12\catcode `\#12\catcode `\^12\catcode `\_12\catcode `\%12\relax}%
	\providecommand \@@startlink[1]{}%
	\providecommand \@@endlink[0]{}%
	\providecommand \url  [0]{\begingroup\@sanitize@url \@url }%
	\providecommand \@url [1]{\endgroup\@href {#1}{\urlprefix }}%
	\providecommand \urlprefix  [0]{URL }%
	\providecommand \Eprint [0]{\href }%
	\providecommand \doibase [0]{https://doi.org/}%
	\providecommand \selectlanguage [0]{\@gobble}%
	\providecommand \bibinfo  [0]{\@secondoftwo}%
	\providecommand \bibfield  [0]{\@secondoftwo}%
	\providecommand \translation [1]{[#1]}%
	\providecommand \BibitemOpen [0]{}%
	\providecommand \bibitemStop [0]{}%
	\providecommand \bibitemNoStop [0]{.\EOS\space}%
	\providecommand \EOS [0]{\spacefactor3000\relax}%
	\providecommand \BibitemShut  [1]{\csname bibitem#1\endcsname}%
	\let\auto@bib@innerbib\@empty
	\bibitem [{\citenamefont {Berendsen}(2007)}]{Berendsen07}%
	\BibitemOpen
	\bibfield  {author} {\bibinfo {author} {\bibfnamefont {H.~J.~C.}\
			\bibnamefont {Berendsen}},\ }\href@noop {} {\emph {\bibinfo {title}
			{Simulating the Physical World}}}\ (\bibinfo  {publisher} {Cambridge
		University Press},\ \bibinfo {address} {Cambridge},\ \bibinfo {year}
	{2007})\BibitemShut {NoStop}%
	\bibitem [{\citenamefont {Peter}\ and\ \citenamefont
		{Kremer}(2009)}]{peter2009}%
	\BibitemOpen
	\bibfield  {author} {\bibinfo {author} {\bibfnamefont {C.}~\bibnamefont
			{Peter}}\ and\ \bibinfo {author} {\bibfnamefont {K.}~\bibnamefont {Kremer}},\
	}\bibfield  {title} {\bibinfo {title} {Multiscale simulation of soft matter
			systems--from the atomistic to the coarse-grained level and back},\ }\href
	{https://doi.org/10.1039/B912027K} {\bibfield  {journal} {\bibinfo  {journal}
			{Soft Matter}\ }\textbf {\bibinfo {volume} {5}},\ \bibinfo {pages} {4357}
		(\bibinfo {year} {2009})}\BibitemShut {NoStop}%
	\bibitem [{\citenamefont {Attinger}\ and\ \citenamefont
		{Koumoutsakos}(2004)}]{Attinger04}%
	\BibitemOpen
	\bibfield  {author} {\bibinfo {author} {\bibfnamefont {S.}~\bibnamefont
			{Attinger}}\ and\ \bibinfo {author} {\bibfnamefont {P.~D.}\ \bibnamefont
			{Koumoutsakos}},\ }\href@noop {} {\emph {\bibinfo {title} {Multiscale
				modelling and simulation}}}\ (\bibinfo  {publisher} {Springer},\ \bibinfo
	{year} {2004})\BibitemShut {NoStop}%
	\bibitem [{\citenamefont {Snook}(2006)}]{snook2006}%
	\BibitemOpen
	\bibfield  {author} {\bibinfo {author} {\bibfnamefont {I.}~\bibnamefont
			{Snook}},\ }\href@noop {} {\emph {\bibinfo {title} {The Langevin and
				generalised Langevin approach to the dynamics of atomic, polymeric and
				colloidal systems}}}\ (\bibinfo  {publisher} {Elsevier},\ \bibinfo {year}
	{2006})\BibitemShut {NoStop}%
	\bibitem [{\citenamefont {Grabert}(2006)}]{grabert2006projection}%
	\BibitemOpen
	\bibfield  {author} {\bibinfo {author} {\bibfnamefont {H.}~\bibnamefont
			{Grabert}},\ }\href {https://books.google.de/books?id=z5t0DgAAQBAJ} {\emph
		{\bibinfo {title} {Projection Operator Techniques in Nonequilibrium
				Statistical Mechanics}}},\ Springer Tracts in Modern Physics\ (\bibinfo
	{publisher} {Springer Berlin Heidelberg},\ \bibinfo {year}
	{2006})\BibitemShut {NoStop}%
	\bibitem [{\citenamefont {Hernandez}\ and\ \citenamefont
		{Somer}(1999)}]{hernandez1999_2}%
	\BibitemOpen
	\bibfield  {author} {\bibinfo {author} {\bibfnamefont {R.}~\bibnamefont
			{Hernandez}}\ and\ \bibinfo {author} {\bibfnamefont {F.}~\bibnamefont
			{Somer}},\ }\bibfield  {title} {\bibinfo {title} {Stochastic dynamics in
			irreversible nonequilibrium environments. 2. a model for thermosetting
			polymerization},\ }\href {https://doi.org/10.1021/jp9836269} {\bibfield
		{journal} {\bibinfo  {journal} {J.~Phys.~Chem.~B}\ }\textbf {\bibinfo
			{volume} {103}},\ \bibinfo {pages} {1070} (\bibinfo {year}
		{1999})}\BibitemShut {NoStop}%
	\bibitem [{\citenamefont {Bhadauria}\ \emph {et~al.}(2015)\citenamefont
		{Bhadauria}, \citenamefont {Sanghi},\ and\ \citenamefont
		{Aluru}}]{Bhadauria2015}%
	\BibitemOpen
	\bibfield  {author} {\bibinfo {author} {\bibfnamefont {R.}~\bibnamefont
			{Bhadauria}}, \bibinfo {author} {\bibfnamefont {T.}~\bibnamefont {Sanghi}},\
		and\ \bibinfo {author} {\bibfnamefont {N.~R.}\ \bibnamefont {Aluru}},\
	}\bibfield  {title} {\bibinfo {title} {Interfacial friction based
			quasi-continuum hydrodynamical model for nanofluidic transport of water},\
	}\href {https://doi.org/10.1063/1.4934678} {\bibfield  {journal} {\bibinfo
			{journal} {The Journal of Chemical Physics}\ }\textbf {\bibinfo {volume}
			{143}},\ \bibinfo {pages} {174702} (\bibinfo {year} {2015})}\BibitemShut
	{NoStop}%
	\bibitem [{\citenamefont {Lei}\ \emph {et~al.}(2016)\citenamefont {Lei},
		\citenamefont {Baker},\ and\ \citenamefont {Li}}]{lei:2016}%
	\BibitemOpen
	\bibfield  {author} {\bibinfo {author} {\bibfnamefont {H.}~\bibnamefont
			{Lei}}, \bibinfo {author} {\bibfnamefont {N.~A.}\ \bibnamefont {Baker}},\
		and\ \bibinfo {author} {\bibfnamefont {X.}~\bibnamefont {Li}},\ }\bibfield
	{title} {\bibinfo {title} {Data-driven parameterization of the generalized
			{L}angevin equation},\ }\href {https://doi.org/10.1073/pnas.1609587113}
	{\bibfield  {journal} {\bibinfo  {journal} {Proc. Natl. Acad. Sci.}\ }\textbf
		{\bibinfo {volume} {113}},\ \bibinfo {pages} {14183} (\bibinfo {year}
		{2016})}\BibitemShut {NoStop}%
	\bibitem [{\citenamefont {Daldrop}\ \emph {et~al.}(2018)\citenamefont
		{Daldrop}, \citenamefont {Kappler}, \citenamefont {Br{\"u}nig},\ and\
		\citenamefont {Netz}}]{Daldrop2018}%
	\BibitemOpen
	\bibfield  {author} {\bibinfo {author} {\bibfnamefont {J.~O.}\ \bibnamefont
			{Daldrop}}, \bibinfo {author} {\bibfnamefont {J.}~\bibnamefont {Kappler}},
		\bibinfo {author} {\bibfnamefont {F.~N.}\ \bibnamefont {Br{\"u}nig}},\ and\
		\bibinfo {author} {\bibfnamefont {R.~R.}\ \bibnamefont {Netz}},\ }\bibfield
	{title} {\bibinfo {title} {Butane dihedral angle dynamics in water is
			dominated by internal friction},\ }\href
	{https://doi.org/10.1073/pnas.1722327115} {\bibfield  {journal} {\bibinfo
			{journal} {Proceedings of the National Academy of Sciences}\ }\textbf
		{\bibinfo {volume} {115}},\ \bibinfo {pages} {5169} (\bibinfo {year}
		{2018})},\ \Eprint
	{https://arxiv.org/abs/https://www.pnas.org/content/115/20/5169.full.pdf}
	{https://www.pnas.org/content/115/20/5169.full.pdf} \BibitemShut {NoStop}%
	\bibitem [{\citenamefont {Wang}\ and\ \citenamefont
		{G{\'o}mez-Bombarelli}(2019)}]{wang2019}%
	\BibitemOpen
	\bibfield  {author} {\bibinfo {author} {\bibfnamefont {W.}~\bibnamefont
			{Wang}}\ and\ \bibinfo {author} {\bibfnamefont {R.}~\bibnamefont
			{G{\'o}mez-Bombarelli}},\ }\bibfield  {title} {\bibinfo {title}
		{Coarse-graining auto-encoders for molecular dynamics},\ }\href
	{https://doi.org/10.1038/s41524-019-0261-5} {\bibfield  {journal} {\bibinfo
			{journal} {npj Computational Materials}\ }\textbf {\bibinfo {volume} {5}},\
		\bibinfo {pages} {1} (\bibinfo {year} {2019})}\BibitemShut {NoStop}%
	\bibitem [{\citenamefont {Kappler}\ \emph {et~al.}(2019)\citenamefont
		{Kappler}, \citenamefont {No\'e},\ and\ \citenamefont {Netz}}]{Kappler2019}%
	\BibitemOpen
	\bibfield  {author} {\bibinfo {author} {\bibfnamefont {J.}~\bibnamefont
			{Kappler}}, \bibinfo {author} {\bibfnamefont {F.}~\bibnamefont {No\'e}},\
		and\ \bibinfo {author} {\bibfnamefont {R.~R.}\ \bibnamefont {Netz}},\
	}\bibfield  {title} {\bibinfo {title} {Cyclization and relaxation dynamics of
			finite-length collapsed self-avoiding polymers},\ }\href
	{https://doi.org/10.1103/PhysRevLett.122.067801} {\bibfield  {journal}
		{\bibinfo  {journal} {Phys. Rev. Lett.}\ }\textbf {\bibinfo {volume} {122}},\
		\bibinfo {pages} {067801} (\bibinfo {year} {2019})}\BibitemShut {NoStop}%
	\bibitem [{\citenamefont {Wang}\ \emph {et~al.}(2020)\citenamefont {Wang},
		\citenamefont {Ma},\ and\ \citenamefont {Pan}}]{wang2020}%
	\BibitemOpen
	\bibfield  {author} {\bibinfo {author} {\bibfnamefont {S.}~\bibnamefont
			{Wang}}, \bibinfo {author} {\bibfnamefont {Z.}~\bibnamefont {Ma}},\ and\
		\bibinfo {author} {\bibfnamefont {W.}~\bibnamefont {Pan}},\ }\bibfield
	{title} {\bibinfo {title} {Data-driven coarse-grained modeling of polymers in
			solution with structural and dynamic properties conserved},\ }\href
	{https://doi.org/10.1039/D0SM01019G} {\bibfield  {journal} {\bibinfo
			{journal} {Soft Matter}\ }\textbf {\bibinfo {volume} {16}},\ \bibinfo {pages}
		{8330} (\bibinfo {year} {2020})}\BibitemShut {NoStop}%
	\bibitem [{\citenamefont {Ozmaian}\ and\ \citenamefont
		{Makarov}(2019)}]{Ozmaian2019}%
	\BibitemOpen
	\bibfield  {author} {\bibinfo {author} {\bibfnamefont {M.}~\bibnamefont
			{Ozmaian}}\ and\ \bibinfo {author} {\bibfnamefont {D.~E.}\ \bibnamefont
			{Makarov}},\ }\bibfield  {title} {\bibinfo {title} {Transition path dynamics
			in the binding of intrinsically disordered proteins: A simulation study},\
	}\bibfield  {journal} {\bibinfo  {journal} {Journal of Chemical Physics}\
	}\textbf {\bibinfo {volume} {151}},\ \href
	{https://doi.org/10.1063/1.5129150} {10.1063/1.5129150} (\bibinfo {year}
	{2019})\BibitemShut {NoStop}%
	\bibitem [{\citenamefont {Grogan}\ \emph {et~al.}(2020)\citenamefont {Grogan},
		\citenamefont {Lei}, \citenamefont {Li},\ and\ \citenamefont
		{Baker}}]{grogan2020}%
	\BibitemOpen
	\bibfield  {author} {\bibinfo {author} {\bibfnamefont {F.}~\bibnamefont
			{Grogan}}, \bibinfo {author} {\bibfnamefont {H.}~\bibnamefont {Lei}},
		\bibinfo {author} {\bibfnamefont {X.}~\bibnamefont {Li}},\ and\ \bibinfo
		{author} {\bibfnamefont {N.~A.}\ \bibnamefont {Baker}},\ }\bibfield  {title}
	{\bibinfo {title} {Data-driven molecular modeling with the generalized
			{Langevin} equation},\ }\href {https://doi.org/10.1016/j.jcp.2020.109633}
	{\bibfield  {journal} {\bibinfo  {journal} {Journal of Computational
				Physics}\ }\textbf {\bibinfo {volume} {418}},\ \bibinfo {pages} {109633}
		(\bibinfo {year} {2020})}\BibitemShut {NoStop}%
	\bibitem [{\citenamefont {Lange}\ and\ \citenamefont
		{Grubmüller}(2006)}]{Lange_2006}%
	\BibitemOpen
	\bibfield  {author} {\bibinfo {author} {\bibfnamefont {O.~F.}\ \bibnamefont
			{Lange}}\ and\ \bibinfo {author} {\bibfnamefont {H.}~\bibnamefont
			{Grubmüller}},\ }\bibfield  {title} {\bibinfo {title} {Collective langevin
			dynamics of conformational motions in proteins},\ }\href
	{https://doi.org/10.1063/1.2199530} {\bibfield  {journal} {\bibinfo
			{journal} {The Journal of Chemical Physics}\ }\textbf {\bibinfo {volume}
			{124}},\ \bibinfo {pages} {214903} (\bibinfo {year} {2006})},\ \Eprint
	{https://arxiv.org/abs/https://doi.org/10.1063/1.2199530}
	{https://doi.org/10.1063/1.2199530} \BibitemShut {NoStop}%
	\bibitem [{\citenamefont {Kinjo}\ and\ \citenamefont
		{Hyodo}(2007)}]{Kinjo_2007}%
	\BibitemOpen
	\bibfield  {author} {\bibinfo {author} {\bibfnamefont {T.}~\bibnamefont
			{Kinjo}}\ and\ \bibinfo {author} {\bibfnamefont {S.-a.}\ \bibnamefont
			{Hyodo}},\ }\bibfield  {title} {\bibinfo {title} {Equation of motion for
			coarse-grained simulation based on microscopic description},\ }\href
	{https://doi.org/10.1103/PhysRevE.75.051109} {\bibfield  {journal} {\bibinfo
			{journal} {Phys. Rev. E}\ }\textbf {\bibinfo {volume} {75}},\ \bibinfo
		{pages} {051109} (\bibinfo {year} {2007})}\BibitemShut {NoStop}%
	\bibitem [{\citenamefont {Chorin}\ \emph {et~al.}(2000)\citenamefont {Chorin},
		\citenamefont {Hald},\ and\ \citenamefont {Kupferman}}]{chorin:2000}%
	\BibitemOpen
	\bibfield  {author} {\bibinfo {author} {\bibfnamefont {A.~J.}\ \bibnamefont
			{Chorin}}, \bibinfo {author} {\bibfnamefont {O.~H.}\ \bibnamefont {Hald}},\
		and\ \bibinfo {author} {\bibfnamefont {R.}~\bibnamefont {Kupferman}},\
	}\bibfield  {title} {\bibinfo {title} {Optimal prediction and the
			{Mori--Zwanzig} representation of irreversible processes},\ }\href
	{https://doi.org/10.1073/pnas.97.7.2968} {\bibfield  {journal} {\bibinfo
			{journal} {Proceedings of the National Academy of Sciences}\ }\textbf
		{\bibinfo {volume} {97}},\ \bibinfo {pages} {2968} (\bibinfo {year}
		{2000})}\BibitemShut {NoStop}%
	\bibitem [{\citenamefont {Chorin}\ \emph {et~al.}(2002)\citenamefont {Chorin},
		\citenamefont {Hald},\ and\ \citenamefont {Kupferman}}]{chorin2002}%
	\BibitemOpen
	\bibfield  {author} {\bibinfo {author} {\bibfnamefont {A.~J.}\ \bibnamefont
			{Chorin}}, \bibinfo {author} {\bibfnamefont {O.~H.}\ \bibnamefont {Hald}},\
		and\ \bibinfo {author} {\bibfnamefont {R.}~\bibnamefont {Kupferman}},\
	}\bibfield  {title} {\bibinfo {title} {Optimal prediction with memory},\
	}\href {https://doi.org/10.1016/S0167-2789(02)00446-3} {\bibfield  {journal}
		{\bibinfo  {journal} {Physica D: Nonlinear Phenomena}\ }\textbf {\bibinfo
			{volume} {166}},\ \bibinfo {pages} {239} (\bibinfo {year}
		{2002})}\BibitemShut {NoStop}%
	\bibitem [{\citenamefont {te~Vrugt}\ and\ \citenamefont
		{Wittkowski}(2020)}]{Vrugt2020}%
	\BibitemOpen
	\bibfield  {author} {\bibinfo {author} {\bibfnamefont {M.}~\bibnamefont
			{te~Vrugt}}\ and\ \bibinfo {author} {\bibfnamefont {R.}~\bibnamefont
			{Wittkowski}},\ }\bibfield  {title} {\bibinfo {title} {Projection operators
			in statistical mechanics: a pedagogical approach},\ }\href
	{https://doi.org/10.1088/1361-6404/ab8e28} {\bibfield  {journal} {\bibinfo
			{journal} {European Journal of Physics}\ }\textbf {\bibinfo {volume} {41}},\
		\bibinfo {pages} {045101} (\bibinfo {year} {2020})}\BibitemShut {NoStop}%
	\bibitem [{\citenamefont {Hij\'on}\ \emph {et~al.}(2010)\citenamefont
		{Hij\'on}, \citenamefont {Espa\~nol}, \citenamefont {Vanden-Eijnden},\ and\
		\citenamefont {Delgado-Buscalioni}}]{Hijon_2010}%
	\BibitemOpen
	\bibfield  {author} {\bibinfo {author} {\bibfnamefont {C.}~\bibnamefont
			{Hij\'on}}, \bibinfo {author} {\bibfnamefont {P.}~\bibnamefont {Espa\~nol}},
		\bibinfo {author} {\bibfnamefont {E.}~\bibnamefont {Vanden-Eijnden}},\ and\
		\bibinfo {author} {\bibfnamefont {R.}~\bibnamefont {Delgado-Buscalioni}},\
	}\bibfield  {title} {\bibinfo {title} {Mori–zwanzig formalism as a
			practical computational tool},\ }\href {https://doi.org/10.1039/B902479B}
	{\bibfield  {journal} {\bibinfo  {journal} {Faraday Discuss.}\ }\textbf
		{\bibinfo {volume} {144}},\ \bibinfo {pages} {301} (\bibinfo {year}
		{2010})}\BibitemShut {NoStop}%
	\bibitem [{\citenamefont {Izvekov}(2017)}]{izvekov2017}%
	\BibitemOpen
	\bibfield  {author} {\bibinfo {author} {\bibfnamefont {S.}~\bibnamefont
			{Izvekov}},\ }\bibfield  {title} {\bibinfo {title} {Microscopic derivation of
			particle-based coarse-grained dynamics: Exact expression for memory
			function},\ }\href {https://doi.org/10.1063/1.4978572} {\bibfield  {journal}
		{\bibinfo  {journal} {The Journal of Chemical Physics}\ }\textbf {\bibinfo
			{volume} {146}},\ \bibinfo {pages} {124109} (\bibinfo {year}
		{2017})}\BibitemShut {NoStop}%
	\bibitem [{\citenamefont {Meyer}\ \emph {et~al.}(2019)\citenamefont {Meyer},
		\citenamefont {Voigtmann},\ and\ \citenamefont
		{Schilling}}]{MeyerVoigtmann_2019}%
	\BibitemOpen
	\bibfield  {author} {\bibinfo {author} {\bibfnamefont {H.}~\bibnamefont
			{Meyer}}, \bibinfo {author} {\bibfnamefont {T.}~\bibnamefont {Voigtmann}},\
		and\ \bibinfo {author} {\bibfnamefont {T.}~\bibnamefont {Schilling}},\
	}\bibfield  {title} {\bibinfo {title} {On the dynamics of reaction
			coordinates in classical, time-dependent, many-body processes},\ }\href
	{https://doi.org/10.1063/1.5090450} {\bibfield  {journal} {\bibinfo
			{journal} {The Journal of Chemical Physics}\ }\textbf {\bibinfo {volume}
			{150}},\ \bibinfo {pages} {174118} (\bibinfo {year} {2019})},\ \Eprint
	{https://arxiv.org/abs/https://doi.org/10.1063/1.5090450}
	{https://doi.org/10.1063/1.5090450} \BibitemShut {NoStop}%
	\bibitem [{\citenamefont {Zwanzig}(2001)}]{zwanzig2001nonequilibrium}%
	\BibitemOpen
	\bibfield  {author} {\bibinfo {author} {\bibfnamefont {R.}~\bibnamefont
			{Zwanzig}},\ }\href@noop {} {\emph {\bibinfo {title} {Nonequilibrium
				Statistical Mechanics}}}\ (\bibinfo  {publisher} {OUP USA},\ \bibinfo {year}
	{2001})\BibitemShut {NoStop}%
	\bibitem [{\citenamefont {Zwanzig}(1960)}]{Zwanzig_60}%
	\BibitemOpen
	\bibfield  {author} {\bibinfo {author} {\bibfnamefont {R.}~\bibnamefont
			{Zwanzig}},\ }\bibfield  {title} {\bibinfo {title} {Ensemble method in the
			theory of irreversibility},\ }\href {https://doi.org/10.1063/1.1731409}
	{\bibfield  {journal} {\bibinfo  {journal} {The Journal of Chemical Physics}\
		}\textbf {\bibinfo {volume} {33}},\ \bibinfo {pages} {1338} (\bibinfo {year}
		{1960})},\ \Eprint {https://arxiv.org/abs/https://doi.org/10.1063/1.1731409}
	{https://doi.org/10.1063/1.1731409} \BibitemShut {NoStop}%
	\bibitem [{\citenamefont {Zwanzig}(1961)}]{Zwanzig_61}%
	\BibitemOpen
	\bibfield  {author} {\bibinfo {author} {\bibfnamefont {R.}~\bibnamefont
			{Zwanzig}},\ }\bibfield  {title} {\bibinfo {title} {Memory effects in
			irreversible thermodynamics},\ }\href
	{https://doi.org/10.1103/PhysRev.124.983} {\bibfield  {journal} {\bibinfo
			{journal} {Phys. Rev.}\ }\textbf {\bibinfo {volume} {124}},\ \bibinfo {pages}
		{983} (\bibinfo {year} {1961})}\BibitemShut {NoStop}%
	\bibitem [{\citenamefont {Mori}(1965)}]{Mori_65}%
	\BibitemOpen
	\bibfield  {author} {\bibinfo {author} {\bibfnamefont {H.}~\bibnamefont
			{Mori}},\ }\bibfield  {title} {\bibinfo {title} {{Transport, Collective
				Motion, and Brownian Motion*)}},\ }\href {https://doi.org/10.1143/PTP.33.423}
	{\bibfield  {journal} {\bibinfo  {journal} {Progress of Theoretical Physics}\
		}\textbf {\bibinfo {volume} {33}},\ \bibinfo {pages} {423} (\bibinfo {year}
		{1965})},\ \Eprint
	{https://arxiv.org/abs/https://academic.oup.com/ptp/article-pdf/33/3/423/5428510/33-3-423.pdf}
	{https://academic.oup.com/ptp/article-pdf/33/3/423/5428510/33-3-423.pdf}
	\BibitemShut {NoStop}%
	\bibitem [{\citenamefont {Kawai}\ and\ \citenamefont
		{Komatsuzaki}(2011)}]{Kawai_2011}%
	\BibitemOpen
	\bibfield  {author} {\bibinfo {author} {\bibfnamefont {S.}~\bibnamefont
			{Kawai}}\ and\ \bibinfo {author} {\bibfnamefont {T.}~\bibnamefont
			{Komatsuzaki}},\ }\bibfield  {title} {\bibinfo {title} {Derivation of the
			generalized langevin equation in nonstationary environments},\ }\href
	{https://doi.org/10.1063/1.3561065} {\bibfield  {journal} {\bibinfo
			{journal} {The Journal of Chemical Physics}\ }\textbf {\bibinfo {volume}
			{134}},\ \bibinfo {pages} {114523} (\bibinfo {year} {2011})},\ \Eprint
	{https://arxiv.org/abs/https://doi.org/10.1063/1.3561065}
	{https://doi.org/10.1063/1.3561065} \BibitemShut {NoStop}%
	\bibitem [{\citenamefont {Izvekov}(2013)}]{Izvekov}%
	\BibitemOpen
	\bibfield  {author} {\bibinfo {author} {\bibfnamefont {S.}~\bibnamefont
			{Izvekov}},\ }\bibfield  {title} {\bibinfo {title} {Microscopic derivation of
			particle-based coarse-grained dynamics},\ }\href
	{https://doi.org/10.1063/1.4795091} {\bibfield  {journal} {\bibinfo
			{journal} {The Journal of Chemical Physics}\ }\textbf {\bibinfo {volume}
			{138}},\ \bibinfo {pages} {134106} (\bibinfo {year} {2013})},\ \Eprint
	{https://arxiv.org/abs/https://doi.org/10.1063/1.4795091}
	{https://doi.org/10.1063/1.4795091} \BibitemShut {NoStop}%
	\bibitem [{\citenamefont {Kauzlarić}\ \emph {et~al.}(2011)\citenamefont
		{Kauzlarić}, \citenamefont {Español}, \citenamefont {Greiner},\ and\
		\citenamefont {Succi}}]{Kauzlaric2011}%
	\BibitemOpen
	\bibfield  {author} {\bibinfo {author} {\bibfnamefont {D.}~\bibnamefont
			{Kauzlarić}}, \bibinfo {author} {\bibfnamefont {P.}~\bibnamefont
			{Español}}, \bibinfo {author} {\bibfnamefont {A.}~\bibnamefont {Greiner}},\
		and\ \bibinfo {author} {\bibfnamefont {S.}~\bibnamefont {Succi}},\ }\bibfield
	{title} {\bibinfo {title} {Three routes to the friction matrix and their
			application to the coarse-graining of atomic lattices},\ }\href
	{https://doi.org/https://doi.org/10.1002/mats.201100014} {\bibfield
		{journal} {\bibinfo  {journal} {Macromolecular Theory and Simulations}\
		}\textbf {\bibinfo {volume} {20}},\ \bibinfo {pages} {526} (\bibinfo {year}
		{2011})}\BibitemShut {NoStop}%
\end{thebibliography}

\end{document}